\documentclass[11pt]{article}
\usepackage{graphicx,array}

\newcommand{\BABARPubYear}    {03}

\newcommand{\BABARProcNumber} {013}
\newcommand{\SLACPubNumber} {10088}
\newcommand{\LANLNumber} {0000}

\input pubboard/babarsym
\newcommand{\etal}      {\mbox{\textsl{et al.}}\xspace}
\newcommand{\forex}     {\mbox{\textsl{e.g.}}\xspace}
\newcommand{\ie}        {\mbox{\textsl{i.e.}}\xspace}
\newcommand{\vs}        {\mbox{\textsl{vs.}}\xspace}
 \def\kg         {\ensuremath{K^{*} \gamma}\xspace}
 \def\xs         {\ensuremath{X_{s} }\xspace}
 \def\xsg        {\ensuremath{\xs \gamma}\xspace}
 \def\pizeta     {\ensuremath{\piz(\eta)}\xspace}
 \def\khiresg    {\ensuremath{K^*(1430) \gamma}\xspace}
 \def\rhog       {\ensuremath{\rho \gamma}\xspace}
 \def\omegag     {\ensuremath{\omega \gamma}\xspace}
 \def\taunu      {\ensuremath{\tau^+ \nut}\xspace}
 \def\munu       {\ensuremath{\mu^+ \num}\xspace}
 \def\knunu      {\ensuremath{K^- \nu \nub}\xspace}
 \def\kll        {\ensuremath{K^{(*)} \ellell}\xspace}
 \def\xsll       {\ensuremath{\xs \ellell}\xspace}
 \def\eg         {\ensuremath{E_{\gamma} }\xspace}

 \def\cth	 {\ensuremath{\cos{\theta_H^*}}\xspace}
 
 \def\mes        {\ensuremath {m_\mathit{ES}}\xspace}
 
 \def\ebeam      {\ensuremath {E^{*}_\mathrm{beam}}\xspace}
 \def\de         {\ensuremath {\Delta E^{*}}\xspace}
 \def\mxs        {\ensuremath{m_{X_{s}} }\xspace}

 \def\lamb       {\ensuremath{\overline{\Lambda}}\xspace}
 \def\mb         {\ensuremath{m_{b} }\xspace}  
 \def\bxsg       {\ensuremath{B \to X_{s} \gamma}\xspace}
 \def\bxdg       {\ensuremath{B \to X_{d} \gamma}\xspace}
 \def\bxsll      {\ensuremath{B \to X_{s} \ellell}\xspace}
 \def\brg        {\ensuremath{B \to \rho \gamma}\xspace}
 \def\brpg       {\ensuremath{\Bp \to \rho^+ \gamma}\xspace}
 \def\brzg       {\ensuremath{\Bz \to \rho^0 \gamma}\xspace}
 \def\bwg        {\ensuremath{\Bz \to \omega \gamma}\xspace}
 \def\bkg        {\ensuremath{B \to \Kstar \gamma}\xspace}

 \def\bkeg       {\ensuremath{\B \to \Kstar(892) \gamma }\xspace}

\def\Journal#1#2#3#4{{#1} {\bf #2}, #3 (#4)}

\def\NIMA{{\em Nucl. Instrum. Methods} A}
\def\NPB{{\em Nucl. Phys.} B}
\def\PLB{{\em Phys. Lett.}  B}
\def\PRL{\em Phys. Rev. Lett.}
\def\PRD{{\em Phys. Rev.} D}
\def\ZPC{{\em Z. Phys.} C}

\def\be{\begin{equation}}
\def\ee{\end{equation}}
\def\bea{\begin{eqnarray}}
\def\eea{\end{eqnarray}}

\long\def\inst#1{\par\nobreak\kern 4pt\nobreak
    {\it #1}\par\vskip 10pt plus 3pt minus 3pt}

\begin{document}
\thispagestyle{empty}

\begin{flushright}
SLAC-PUB-\SLACPubNumber \\
\babar-PROC-\BABARPubYear/\BABARProcNumber \\
hep-ex/\LANLNumber \\
August, 2003 \\
\end{flushright}

\par\vskip 4cm

\begin{center}
{\Large \bf Rare Radiative and Leptonic \B Decays:  Recent Results
from \babar\ and Belle}
\end{center}
\bigskip

\begin{center}
\large 
A.M. Eisner\\
University of California at Santa Cruz, Institute for Particle Physics \\
1156 High Street, Santa Cruz, CA 95064, USA \\
\end{center}
\bigskip \bigskip

\begin{center}
\large \bf Abstract
\end{center}
Recent \babar\ and Belle results on rare \B decays 
involving photons or lepton pairs are reviewed.  New preliminary limits from
both experiments for \brg and \bwg, and from \babar\ 
for $\Bm\to\knunu$ and $\Bp\to\taunu$ are presented.  
Also summarized from the past year are Belle measurements of 
\bxsll and $\Bz\to K_2^*(1430)^0\gamma$, preliminary 
\babar\ measurements of \bxsg, and preliminary results from 
\babar\ and Belle on $B\to\kll$.  No significant deviations from 
Standard Model predictions have yet been found.

\vfill
\begin{center}
Contributed to the Proceedings of the XXXVIII Rencontres de Moriond,
Electroweak Interactions and Unified Theories, \\
3/15/2003---3/22/2003, Les Arcs, France
\end{center}

\vspace{1.0cm}
\begin{center}
{\em Stanford Linear Accelerator Center, Stanford University, 
Stanford, CA 94309} \\ \vspace{0.1cm}\hrule\vspace{0.1cm}
Work supported in part by Department of Energy contracts DE-AC03-76SF00515
and DE-FG03-92ER40689.
\end{center}


\section{Introduction}

This review covers results from \babar\ and Belle since the 2002
Moriond-EW conference in the area of rare \B decays
involving photons or lepton pairs.  In particular, the final
states to be presented are \rhog and \omegag (new results from
both experiments), \khiresg (Belle, published), \xsg (\babar,
of ICHEP-2002 vintage), \kll (both experiments, ICHEP-2002),
\xsll (Belle, published), \knunu (new result from \babar)
and \taunu (new results from \babar).  Here \xs refers to inclusvie strange
hadronic states, and \ellell to either a di-muon or di-electron pair.

\begin{figure}[htb]
 \begin{minipage}[c]{0.34\textwidth}
  \begin{center}
   \includegraphics[width=\textwidth,clip]{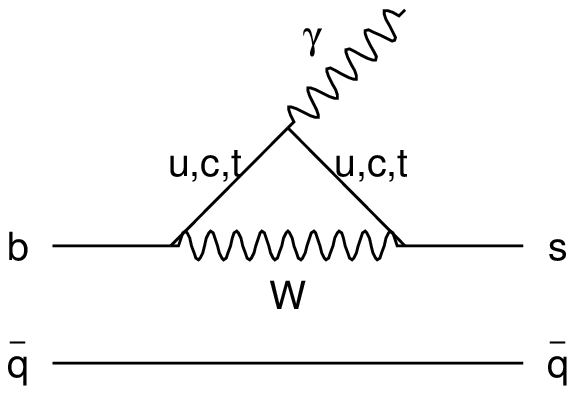}
  \end{center}
 \end{minipage}
 \begin{minipage}[c]{0.04\textwidth}
  \mbox{}
 \end{minipage}
 \begin{minipage}[c]{0.59\textwidth}
  \begin{center}
   \includegraphics[width=\textwidth,clip]{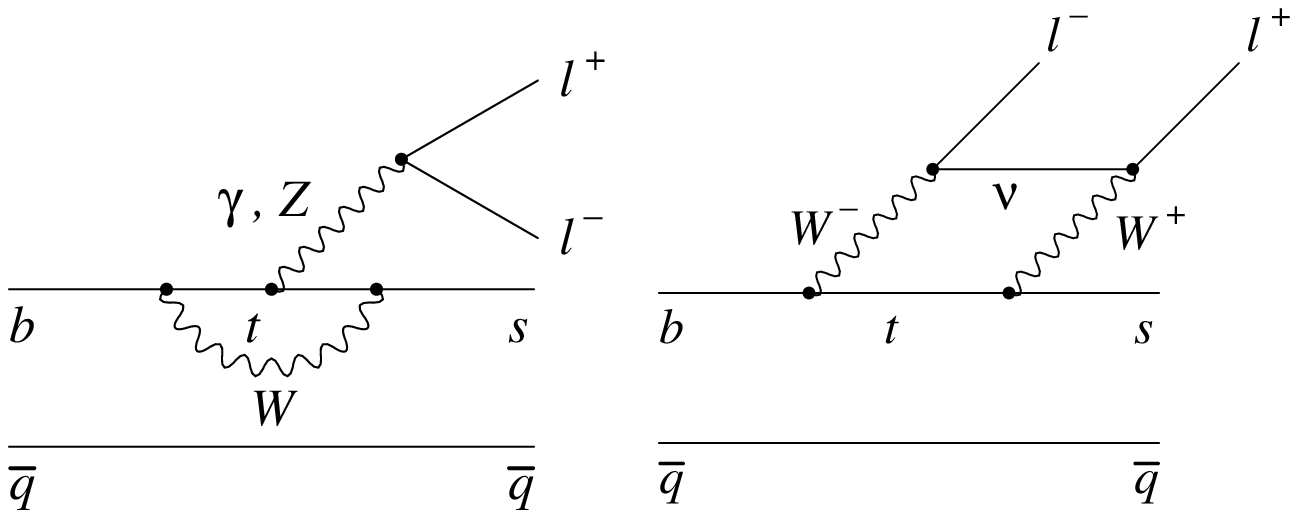}
  \end{center}
 \end{minipage} \\
 \begin{minipage}[b]{0.34\textwidth}
  \caption{\bxsg in the Standard Model (penguin diagram)\label{fig:bxsg_diag}}
 \end{minipage}
 \begin{minipage}[b]{0.04\textwidth}
  \mbox{}
 \end{minipage}
 \begin{minipage}[b]{0.59\textwidth}
  \caption{\bxsll in the Standard Model (penguin and box diagrams)
  \label{fig:xll_diag}}
 \end{minipage}
\end{figure}

None of these decays occurs at tree level; rather, all involve internal
loops or boxes or (for the purely leptonic \taunu decay) $u\overline b$
annihilation.  In all cases the predicted branching fractions are low,
and in most the theoretical uncertainties in the Standard Model (``SM'') are 
low.  (Because of fragmentation effects, uncertainties are larger whenever 
a specific hadron appears in the final state.)
Thus these are good places to look for non-standard contributions,
with the potential for new particles showing up virtually in the loops.
For example, Figure~\ref{fig:bxsg_diag} illustrates one of the two 
``radiative penguin'' diagrams responsible for \bxsg in the Standard Model, 
while Figure~\ref{fig:xll_diag} shows penguin and box diagrams for \bxsll.
Of course, there are QCD corrections to these lowest-order amplitudes.

\section{Radiative $\mathbf{B}$ Decays}

Next-to-leading order (NLO) predictions for \bxsg in the Standard Model 
have been made to about 10\% precision\,\cite{bib:Gambino,bib:Buras}.
The easiest component of this to measure is the exclusive decay
\bkeg, which has a branching fraction of $\sim 4\times 10^{-5}$, and
has been determined in each charge state to about 10\% 
precision\,\cite{bib:PDG}.  Unfortunately, predictions are much less
certain, since hadronization is involved.

This review presents results which extend the experimental
knowledge in several directions:  to higher \Kstar resonances, to
the inclusive \bxsg process, and to searches for \bxdg.
The latter proceeds in the Standard Model via a diagram analogous
to Figure~\ref{fig:bxsg_diag}, with the \s quark replaced by a \d.
Because the amplitude is dominated by a \t quark in the loop, the
\bxdg yield is reduced by 
$\sim (\Vtd/\Vts)^2 \sim 0.04$ compared to \bxsg.  As with \bxsg,
the first searches are for exclusive states, the $\rho$ and the
$\omega$.  

\subsection{Searches for \brpg, \brzg and \bwg}\label{sec:rho}

Predictions for \BR(\brpg) in the Standard 
Model\,\cite{bib:Ali_Parkhomenko,bib:Bosch} 
are $\sim 0.9\ \mathrm{to}\ 1.5 \times 10^{-6}$.
Using isospin symmetry and the quark model, one expects 
$\BR(\brzg)\ \ \mathrm{and}\ \  
 \BR(\bwg)\ \ \mathrm{each}\ \approx \BR(\brpg)/2$.
\babar\,\cite{bib:babar_rho} and Belle\,\cite{bib:belle_rho} are reporting
updated searches for these three decays.  

Backgrounds for these measurements are:
\begin{itemize}
 \item[$\bullet$] Continuum (\uubar, \ddbar, \ssbar, \ccbar) events with the
  high-energy photon mostly from \pizeta decay or initial state
  radiation (``ISR'').  These events have a jetlike topology.
 \item[$\bullet$] For \brg: \bkg with a $K$ misidentified as a $\pi$.  This 
  needs to be suppressed, because \BR(\bkg) is at least 25 times larger
  than the \BR\ expected for signal.
 \item[$\bullet$] Mainly for \brpg:  a small contribution from 
  $\B\to\rho(\omega)\piz$.
\end{itemize}
\babar\ (Belle) combines topological quantities into a neural net (Fisher 
discriminant) to reduce continuum background.  Particle-identification (PID)
and kinematic criteria are used to suppress \bkg.   In particular, \babar\
achieves very good rejection using the DIRC\,\,\cite{bib:BABAR} plus dE/dx;
as illustrated in their paper\,~\cite{bib:babar_rho}, the \babar\ $K$-to-$\pi$
misidentification probability in nearly all of the relevant region is
determined from control samples to be less than 1\%.

Results for exclusive \B decays are typically presented using the
following kinematic variables.  If $(E_B^*, \vec{p}_B^*) $ 
is the four-momentum of a reconstructed \B candidate in the overall 
CM (\Y4S) frame, we define
\be
 \de \equiv E_B^* - \ebeam \ ,
 \label{eq:de}
\ee
\be
 \mes\ (\mathrm{or}\ M_{bc}) \equiv \sqrt{{\ebeam}^2 - p_B^{*2}}\ .
 \label{eq:mes}
\ee
The latter is called the \emph{energy-substituted} (\babar) or
\emph{beam-constrained} (Belle) mass.  Signal events peak at \de near 0 
and \mes near $M_B$ (above $\approx 5.27\gevcc$); whereas continuum
background lacks peaks, and is usually
fit in \mes by a single-parameter ``ARGUS'' threshold 
function\,\cite{bib:argus}.

\begin{figure}[htb]
 \centering
 \includegraphics[width=0.5\textwidth,clip]{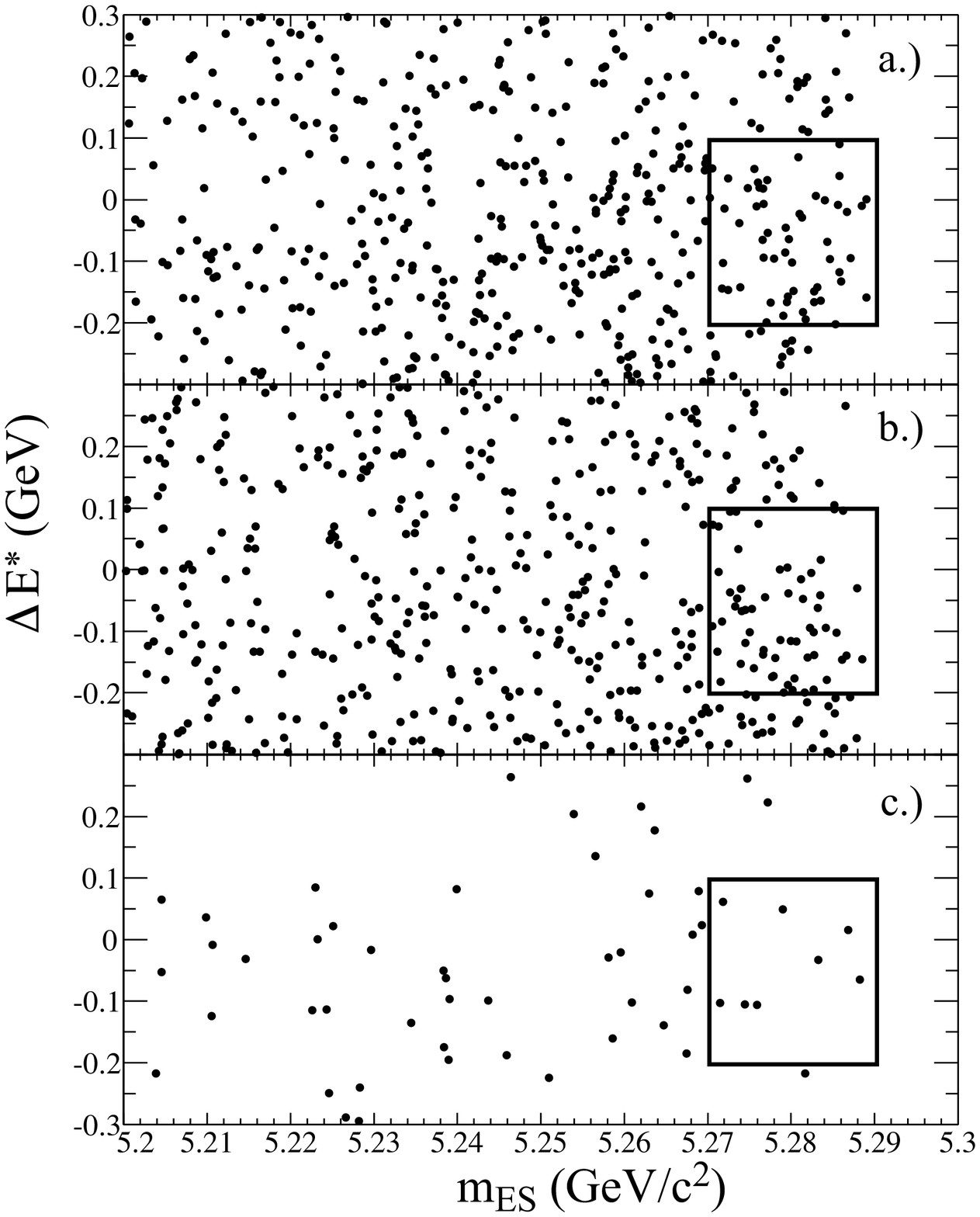}
 \caption{\de \vs \mes for \babar\ data: (a) \brzg, (b) \brpg,
  (c) \bwg candidates after all selection cuts.  
  For $\BR(\brpg) = 1.0 \times 10^{-6}$ (and the others half as large),
  one would expect 5.0, 6.0 and 1.7 signal events in the three signal boxes,
  compared to \ 49, \ 72, and \ 9 background events, respectively.}
  \label{fig:babar_rho_omega}.
\end{figure}

Signal extraction in \babar\ uses an unbinned maximum likelihood fit in
\mes, \de and (for \rhog) $m_{\pi\pi}$.  As is standard in most such
\babar\ analysis, a signal region is ``blinded'' (not looked at)
until all cuts and procedures are finalized.  Figure~\ref{fig:babar_rho_omega}
shows the \babar\ data.

Neither \babar\,\cite{bib:babar_rho} nor Belle\,\cite{bib:belle_rho} sees
any evidence for a signal.  Table~\ref{tab:rho_omega} shows upper limits, 
along with those from an earlier search.   The Belle and \babar\ limits
allow for systematic uncertainties, using a method based on
Cousins and Highland\,\cite{bib:Cousins}.

\begin{table}[htb]
 \begin{center}
  \vspace{-0.1in}
  \caption{Upper limits for \brg and \bwg. \label{tab:rho_omega}}
  \vspace{0.15in}
  \renewcommand{\arraystretch}{1.2}
  \addtolength{\extrarowheight}{3pt}
  \begin{tabular}{|l|c|c|c|c|}
   \hline
   & $\int\mathcal{L}dt$ & \multicolumn{3}{c|}{90\% CL Upper Limits
    on \BR}\\
   \cline{3-5}
   Experiment & (\invfb) & \brzg & \brpg & \bwg \\ \hline
   CLEO-II\,\cite{bib:cleo_rho}   & 9.2& $17\times 10^{-6} $& $13\times 10^{-6} $& $9.2\times 10^{-6}$ \\
   Belle\,\cite{bib:belle_rho} & 78 & $2.6\times 10^{-6}$& $2.7\times 10^{-6}$& $4.4\times 10^{-6}$ \\
   \babar\,\cite{bib:babar_rho}& 78 & $1.2\times 10^{-6}$& $2.1\times 10^{-6}$& $1.0\times 10^{-6}$ \\
   \hline
  \end{tabular}
 \end{center}
\end{table}

Belle quotes a combined limit on
$\BR(\B\to (\rho+\omega)\gamma) \equiv \BR(\brpg) = 2\BR(\brzg)
= 2\BR(\bwg)$ of $3.0\times 10^{-6}$ (90\% CL).  Their 5-mode
analysis (including two \kg charge states) then implies
$\BR(\B\to (\rho+\omega)\gamma)/\BR(\bkg) < 0.081$ (90\% CL).

\babar\ combines just the two $\rho$ modes, finding
$\BR(\brg) \equiv \BR(\brpg) = 2\BR(\brzg) < 1.9\times 10^{-6}$ (90\% CL),
which is just above the high end of the range of Standard Model predictions.
With the published \babar\ value for \bkg, this implies 
$\BR(\brg)/\BR(\bkg) < 0.047$ (90\% CL).

\subsection{Higher $K^*$ Resonances}\label{sec:higher}

\begin{figure}[htb]
 \begin{center}
  \includegraphics[width=0.5\textwidth,clip]{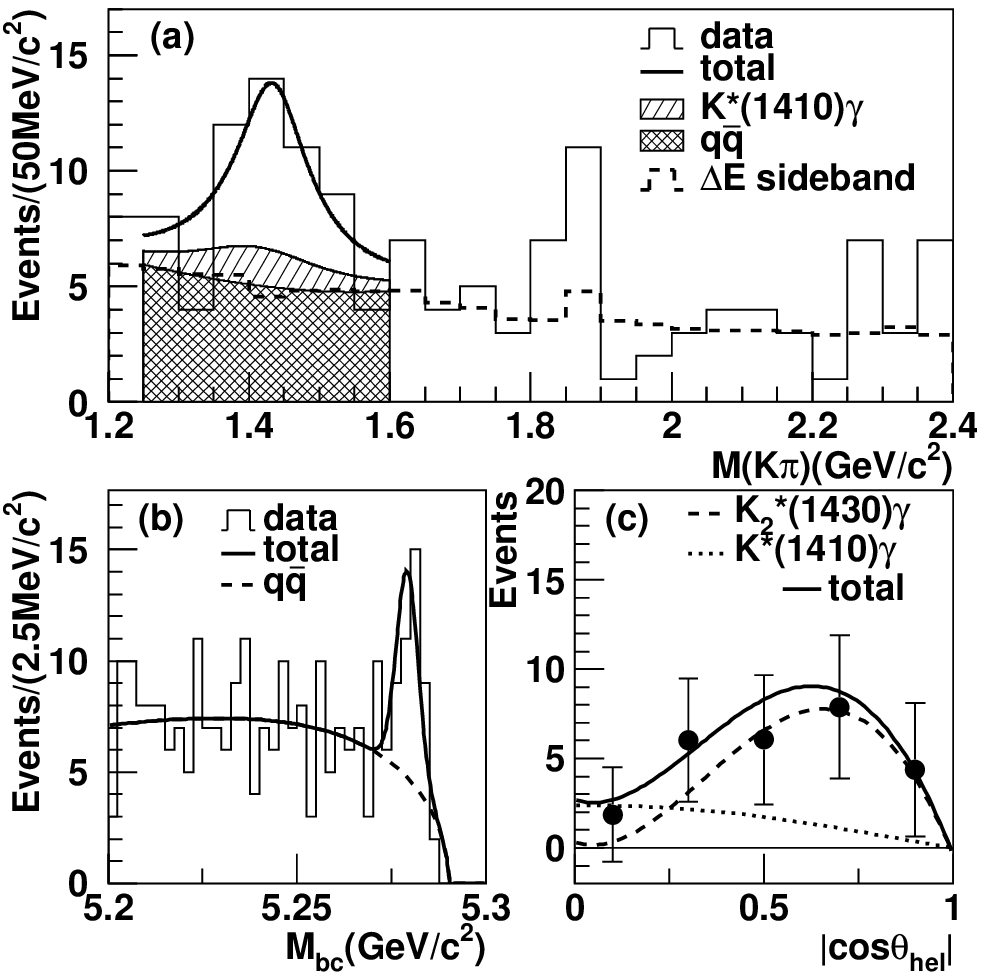} 
  \caption{Distribution of Belle $B\to K^+\pi^-\gamma$ candidates:
   (a) $M_{K\pi}$ (for $M_{bc} > 5.27\gevcc$), (b) $M_{bc}$ (for
   $1.25 < M_{K\pi} < 1.6\gevcc$) and (c) \cth (signal only, with
   $M_{bc} > 5.27\gevcc$).  The curves in (a) and (c) are results of an 
   unbinned maximum likelihood fit.  \label{fig:Kpigamma}}
 \end{center}
\end{figure}

Belle has published measurements of $K\pi\gamma$ and
$K\pi\pi\gamma$ \B decays\,\cite{bib:higher} using 29.4\invfb of data. 
After continuum suppression (topological) cuts and a \de cut, 
the $K\pi\gamma$ signal peak in $M_{bc}$ was fitted to a sum of 
$K_2^*(1430)$, $K^*(1410)$ and a non-resonant contribution, 
using $M_{K\pi}$ and the decay helicity angle \cth to distinguish them.
Figure~\ref{fig:Kpigamma} shows the results.

Only the $K_2^*(1430)$ component was significant, with the fit implying  
$\BR(\Bz\rightarrow K_2^*(1430)^0\gamma) = 
(1.3\pm 0.5 \pm 0.1)\times 10^{-5}\ .$  Predictions cover a wide range;
this result is consistent with those from a relativistic form factor 
model\,\cite{bib:Veseli}.

Belle also found $\BR(\Bp\rightarrow K^+\pi^-\pi^+\gamma) = 
(2.4\pm 0.5{\:}^{+0.4}_{-0.2})\times 10^{-5}$, dominated by 
$K^{*0}\pi^+$ and $K^+\rho^0$.  

\subsection{Inclusive \bxsg}\label{sec:bxsg}

Inclusive \bxsg measurements are of interest for several reasons:

\begin{itemize}
 \item[$\bullet$] The NLO Standard Model computation of the branching 
  fraction\,\cite{bib:Gambino,bib:Buras} has precision below 10\%,
  much better than for exclusive channels.  A measurement of
  \BR(\bxsg) is sensitive to new physics.
 \item[$\bullet$] The \eg spectrum has been parameterized\,\cite{bib:Kagan} 
  in terms of just the \b quark mass (\mb) and a Fermi momentum
  parameter ($\lambda_1$).  This spectrum is largely insensitive to
  new physics.
 \item[$\bullet$] The spectrum and especially its moments are related 
  to those in $B\rightarrow X\ell\nu$, and hence relevant to the extraction
  of \Vcb and \Vub in semileptonic \B decay.
\end{itemize}

Two independent preliminary \babar\ measurements of \bxsg,
were presented at ICHEP-2002 -- 
one fully inclusive\,\cite{bib:inclbsg_babar},
the other using a semi-inclusive sum of exclusive 
modes\,\cite{bib:semibsg_babar}.
Because there have been no more recent
results, I just report the overall situation as of ICHEP-2002.

\begin{figure}[htb]
 \begin{minipage}[c]{0.49\textwidth}
  \begin{center}
   \includegraphics[width=\textwidth,clip]{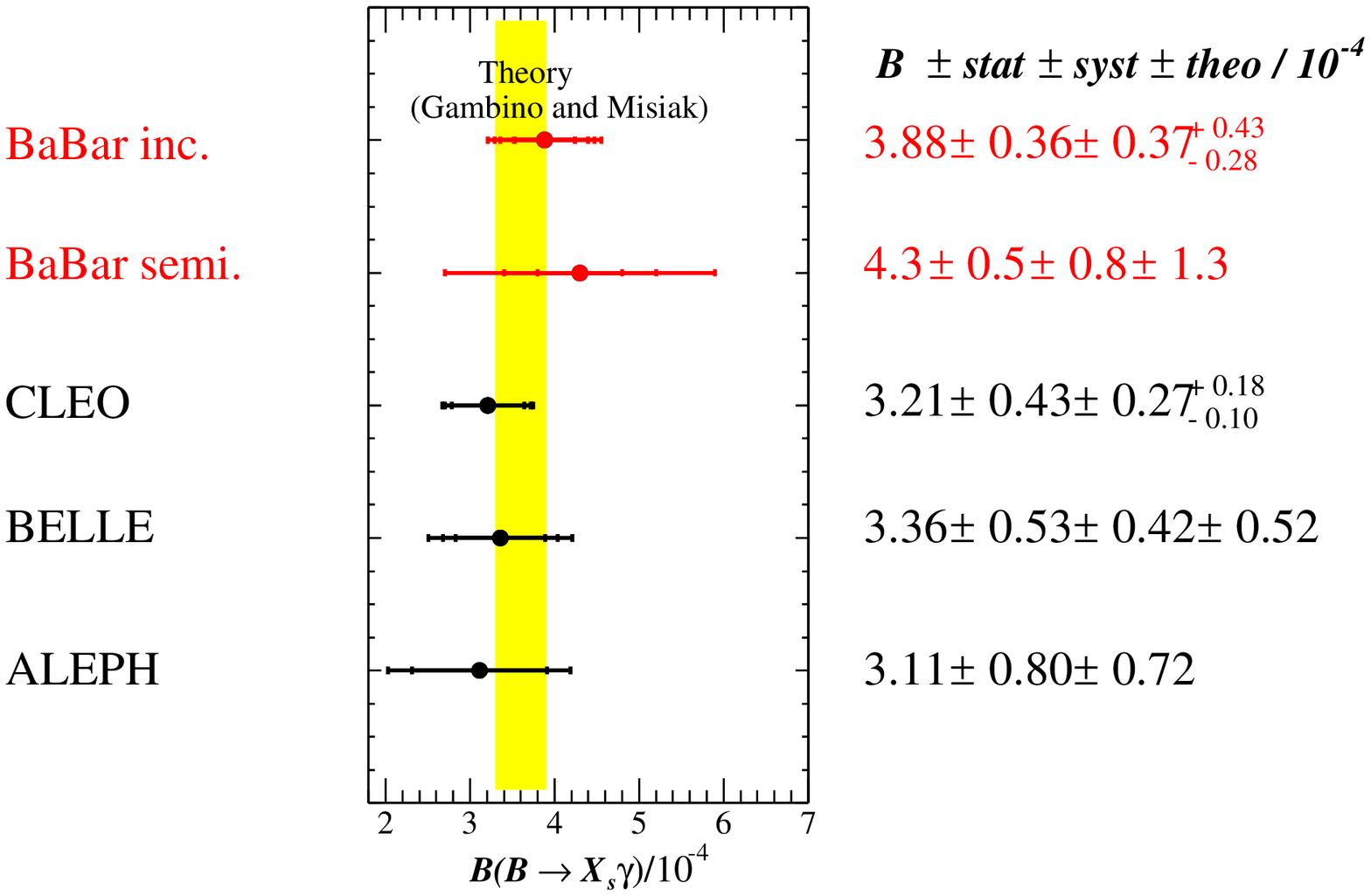}
  \end{center}
 \end{minipage}
 \begin{minipage}[c]{0.49\textwidth}
  \begin{center}
   \includegraphics[width=\textwidth,clip]{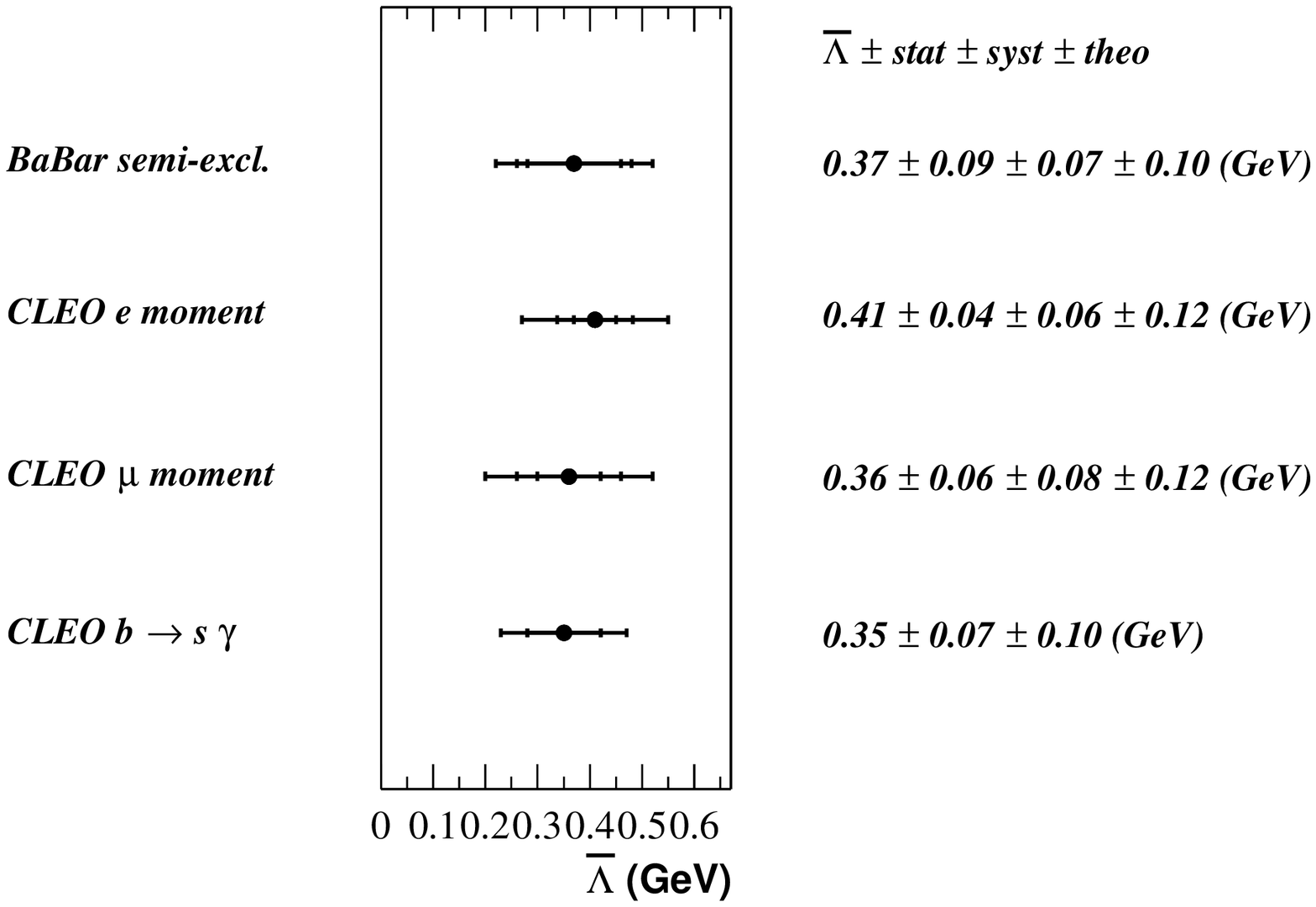}
  \end{center}
 \end{minipage} \\[-0.1in]
 \begin{minipage}[b]{0.47\textwidth}
  \caption{Measurements of \BR(\bxsg) compared to the NLO Standard
   Model prediction (shaded band).  
   See ``theo'' explanation and caveats in text.
   \label{fig:inclbsg_BR}}
 \end{minipage}
 \begin{minipage}[b]{0.04\textwidth}
  \mbox{}
 \end{minipage}
 \begin{minipage}[b]{0.47\textwidth}
  \caption{Measurements of \lamb in HQET (in GeV) from \bxsg and 
   semi-leptonic \B decays.
   \label{fig:LambdaBar}}
 \end{minipage}
\end{figure}

Figure~\ref{fig:inclbsg_BR} summarizes the branching fraction measurements,
most with a model-dependence uncertainty, denoted ``$\pm$ theo''.
Experiments differ in the minimum value of \eg used for analysis
(a consequence of differences in background severity).  Thus
extrapolations to lower \eg (higher $m_{X_s}$), using the
Kagan-Neubert (KN) model\,\cite{bib:Kagan}, have differing 
model-dependence uncertainty, largely from the choice of \mb.
The \babar\ semi-inclusive measurement fits the measured hadronic mass
spectrum rather than extrapolating, but its model-dependence error is
likewise due to limited knowledge of the KN parameters.
[Two caveats at the less-than-3\% level:  first, there is a lack of
standardization of the point extrapolated \emph{to}; second, the KN model 
may have omitted a small radiative effect\,\cite{bib:Bigi} in the 
extrapolation.]
Results to date are in agreement with the Standard Model prediction.

The mean $<\eg>$ above the measured \eg cutoff can be related to
\lamb in HQET, using a procedure from Ligeti \etal\cite{bib:Ligeti}.
Figure~\ref{fig:LambdaBar} summarizes values of \lamb obtained
from either \bxsg or semi-leptonic decays.  Uncertainties are
still large.

\section{Decays Resulting in Charged-Lepton Pairs}

Detecting a lepton pair (\epem or \mumu) instead of a photon
opens up a new dimension (the lepton pair mass), and adds the box 
and ``$Z$ penguin'' amplitudes (Fig.~\ref{fig:xll_diag}),
providing more ways in which new physics might enter.  \bxsg is
sensitive mainly to the magnitude of the Wilson coefficient $C_7$
in the Operator Product Expansion, 
while \bxsll is sensitive also to its phase and to coefficients $C_9$
and $C_{10}$.
The penalty one pays is an expected branching fraction 50 or so times 
smaller in the Standard Model.
As with \bxsg, experimenters began with exclusive modes, but
have already moved on to the inclusive measurement.

\subsection{$\B\rightarrow\kll$}\label{sec:kll}

\begin{figure}[htb]
 \begin{minipage}[c]{0.48\textwidth}
  \begin{center}
   \includegraphics[width=\textwidth,clip]{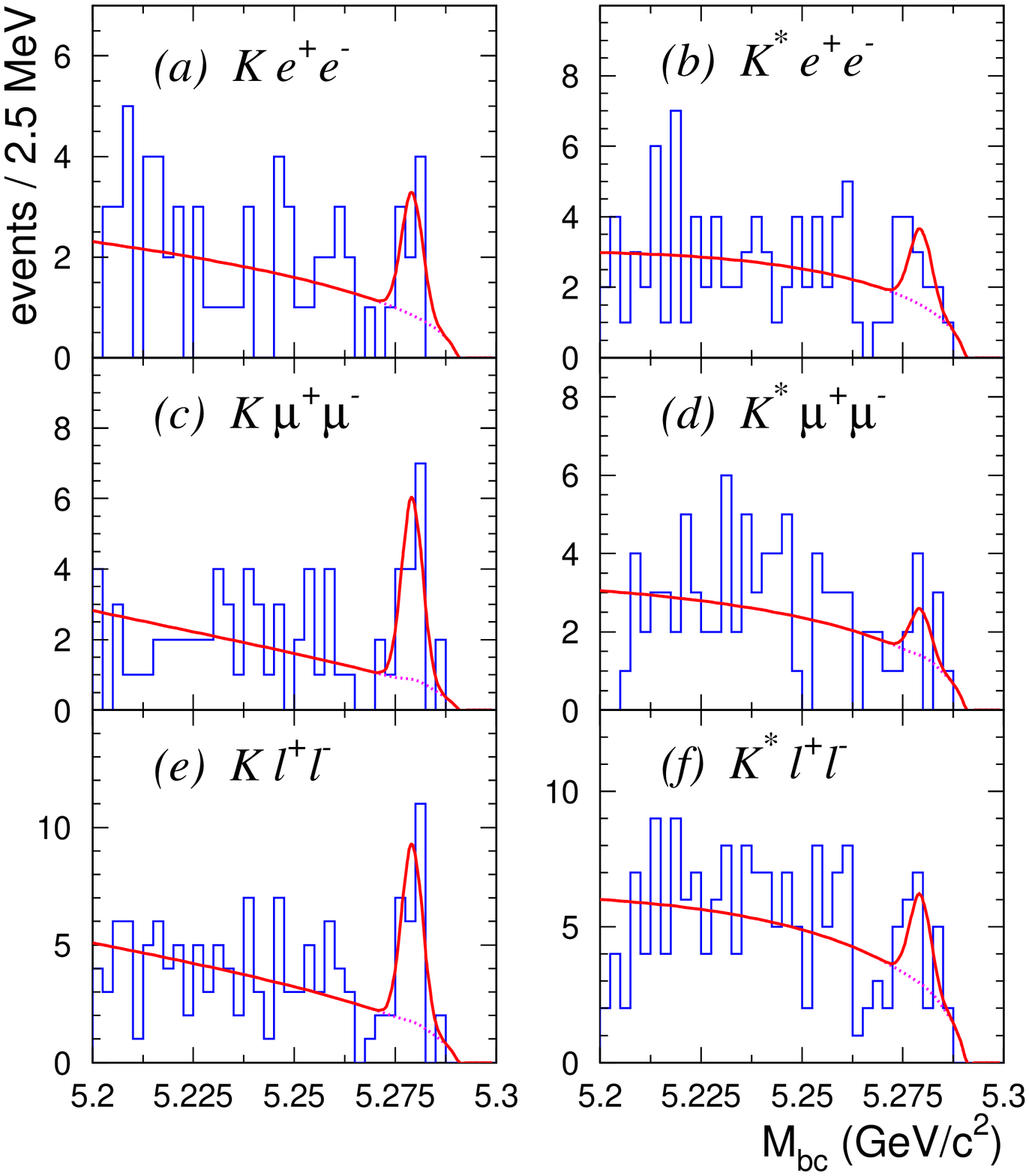}
  \end{center}
 \end{minipage}
 \begin{minipage}[c]{0.03\textwidth}
  \mbox{}
 \end{minipage}
 \begin{minipage}[c]{0.48\textwidth}
  \begin{center}
   \includegraphics[width=\textwidth,clip]{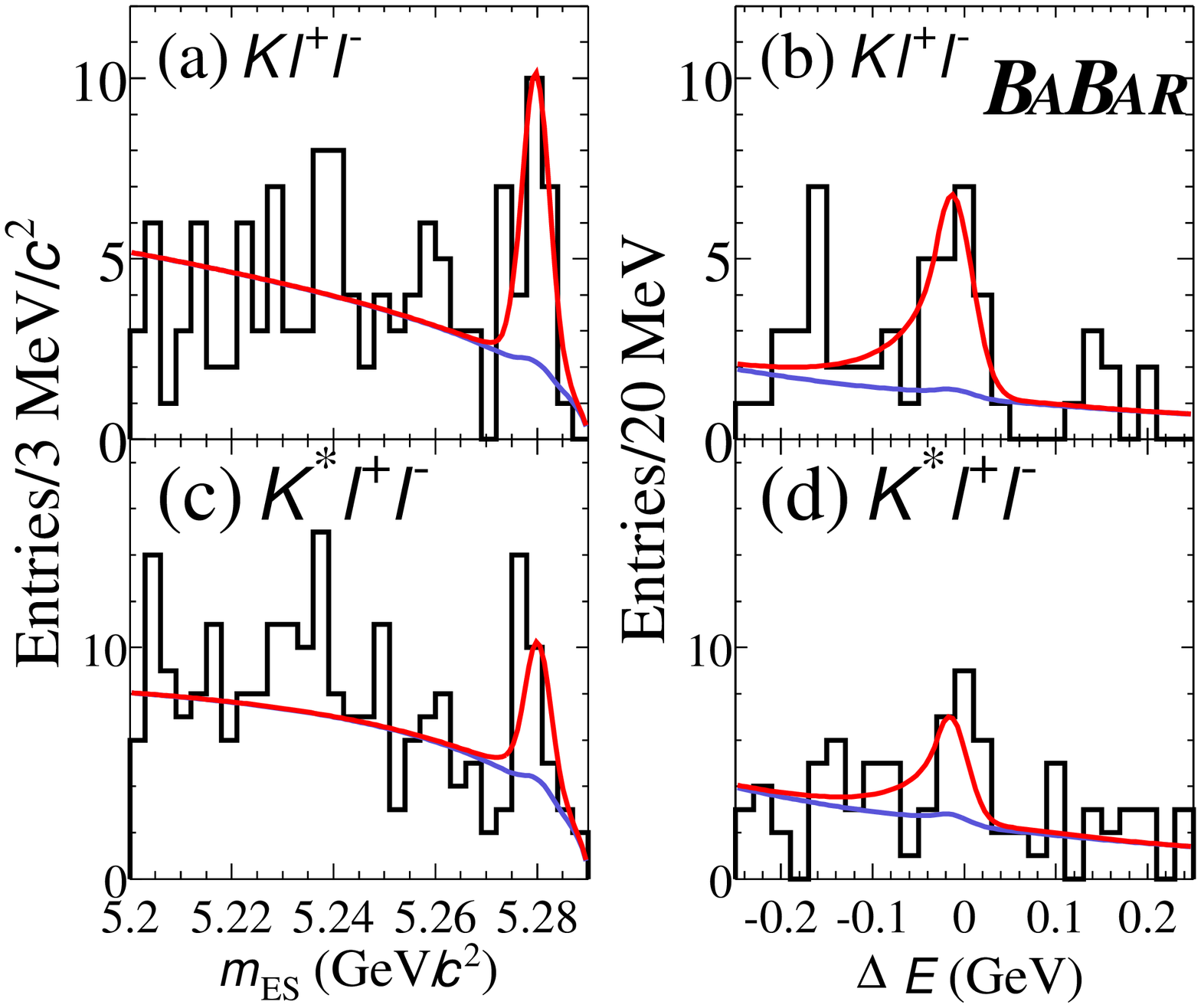}
  \end{center}
 \end{minipage} \\[-0.2in]
 \begin{minipage}[b]{0.48\textwidth}
  \caption{Belle preliminary spectra (ICHEP-2002, 60.1\invfb) for $\B\to\kll$ 
   after \de cuts, summed over charge states.  The spectra in (e) and (f)
   are the sums of those above.  The curves show binned maximum 
   likelihood fits to each $M_{bc}$ spectrum.
   \label{fig:belle_kll}}
 \end{minipage}
 \begin{minipage}[b]{0.03\textwidth}
  \mbox{}
 \end{minipage}
 \begin{minipage}[b]{0.48\textwidth}
  \caption{\babar\ preliminary spectra (ICHEP-2002, 77.8\invfb) for 
   $\B\to\kll$ summed over charge states and lepton flavor.  The curves
   are projections of unbinned two-dimensional maximum likelihood fits to 
   \mes and \de, with relative branching fractions for the 
   four component processes constrained (see text).
   \label{fig:babar_kll}}
 \end{minipage}
\end{figure}

Belle\,\cite{bib:firstKll} first established a signal for $K\ellell$
from 29\invfb.  The most recent Belle\,\cite{bib:belle_Kll} 
and \babar\,\cite{bib:babar_Kll} results were presented at
ICHEP in 2002.  Eight modes are measured:  
\kaon/\Kstar, charged/neutral, \epem/\mumu.  
Both experiments veto backgrounds consistent with 
$\jpsi (\psi') \rightarrow \ellell$ (allowing also for Bremsstrahlung 
in the case of \epem) and suppress continuum and combinatoric \BB 
backgrounds with topological and kinematic cuts.
They then fit to separate the signal from the residual background.
Spectra and fits are shown in
Figures~\ref{fig:belle_kll} and~\ref{fig:babar_kll}.

The \Kp and \Kz modes are combined by assuming isospin symmetry.
Lepton universality is assumed for $\kaon\ellell$.  But $\kaon^*\ellell$
has a photon pole at $m^2_{\ellell} = 0$, resulting in 
$\BR(\B\rightarrow\Kstar\epem)/\BR(\B\rightarrow\Kstar\mumu) > 1$.
Belle constrains this ratio to 1.33, from Ali \etal\,\cite{bib:Ali_2002};
while \babar\ used a ratio of 1.21 from an earlier paper\,\cite{bib:Ali_2000}.
Results are shown in Table~\ref{tab:kll}.
The NNLO Standard Model prediction from Ali \etal\cite{bib:Ali_2002} is 
$\BR(B\rightarrow K\ellell) = (0.35\pm 0.13)\times 10^{-6}$, more than
$1\sigma$ below each measurement, but there \emph{are} higher predictions
in the literature, and no significant disagreement can be claimed.

\begin{table}[p]
 \begin{center}
  \vspace{-0.1in}
  \caption{Belle and \babar\ preliminary (ICHEP-2002) results for
   $\B\rightarrow\kll$.  Note: the Belle limit for $\kaon^*\ellell$
   is an average of \epem and \mumu modes, while the \babar\ limit is
   scaled to \epem.  If the \babar\ $\kaon^*\ellell$ result were 
   interpreted as a signal, its
   value would be $(1.68^{\,+0.68}_{\,-0.58} \pm 0.28)\times 10^{-6}$, 
   but its significance is only $2.8\sigma$.
   \label{tab:kll}}
  \vspace{0.1in}
  \renewcommand{\arraystretch}{1.2}
  \addtolength{\extrarowheight}{3pt}
  \begin{tabular}{|l|c|c|}  
   \hline
   Branching Ratios & Belle $\times 10^6$ & \babar\ $\times 10^6$ \\
   \hline
   $\B\rightarrow K\mumu$       & $0.80^{\,+0.28}_{\,-0.23} \pm 0.09$ & \\
   $\B\rightarrow K\ellell$     & $0.58^{\,+0.17}_{\,-0.15} \pm 0.06$ &
                       $0.78^{\,+0.24}_{\,-0.20}{}^{\,+0.11}_{\,-0.18}$ \\
   \hline
   $\B\rightarrow\Kstar\ellell$ & $ <1.4$ (90\% CL) &
                                  $ <3.0$ (90\% CL)                     \\
   \hline
  \end{tabular}
 \end{center}
\end{table}

\subsection{A First Measurement of \bxsll}\label{sec:bxsll}

As with \bxsg, \bxsll can be much more precisely computed in the 
Standard Model than can its exclusive counterparts.  Belle has now published
a first measurement\,\cite{bib:belle_xsll} for dilepton masses above
0.2\gevcc.  The NNLO prediction of Ali \etal\cite{bib:Ali_2002} is
$\BR(\bxsll \mathrm{with} M_{\ellell} > 0.2\gevcc) = 
(4.2\pm 0.7)\times 10^{-6}$.

Belle's measurement is semi-inclusive, using 18 specific final states
(one \Kp or \KS, 0 to 4 $\pi$, including 0 or 1 \piz).  In order to
estimate signal detection efficiencies and optimize selection cuts,
the Monte Carlo detector simulation uses theoretical models as input for the
dilepton mass spectrum\,\cite{bib:Ali_2002} and the hadronic mass ($M_{X_s}$)
Fermi spectrum\,\cite{bib:Ali_1997}, while final hadronic states are
produced according to JETSET.  For $M_{X_s} < 1.1\gevcc$, the exclusive-state
predictions\,\cite{bib:Ali_2002} are used.
 
\begin{figure}[p]
 \begin{minipage}[c]{0.48\textwidth}
  \begin{center}
   \includegraphics[width=\textwidth,clip]{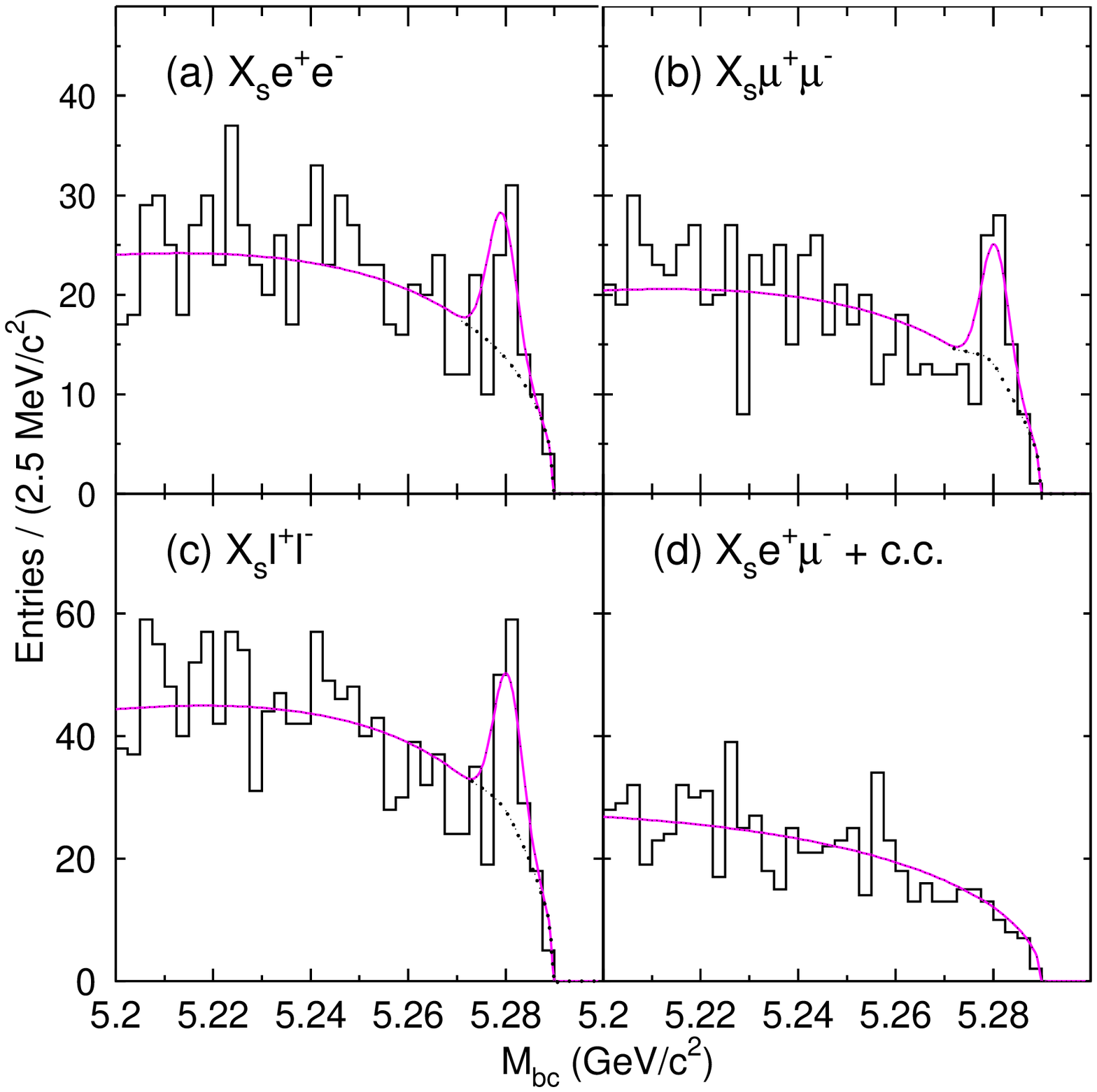}
  \end{center}
 \end{minipage}
 \begin{minipage}[c]{0.03\textwidth}
  \mbox{}
 \end{minipage}
 \begin{minipage}[c]{0.48\textwidth}
  \begin{center}
   \includegraphics[width=\textwidth,clip]{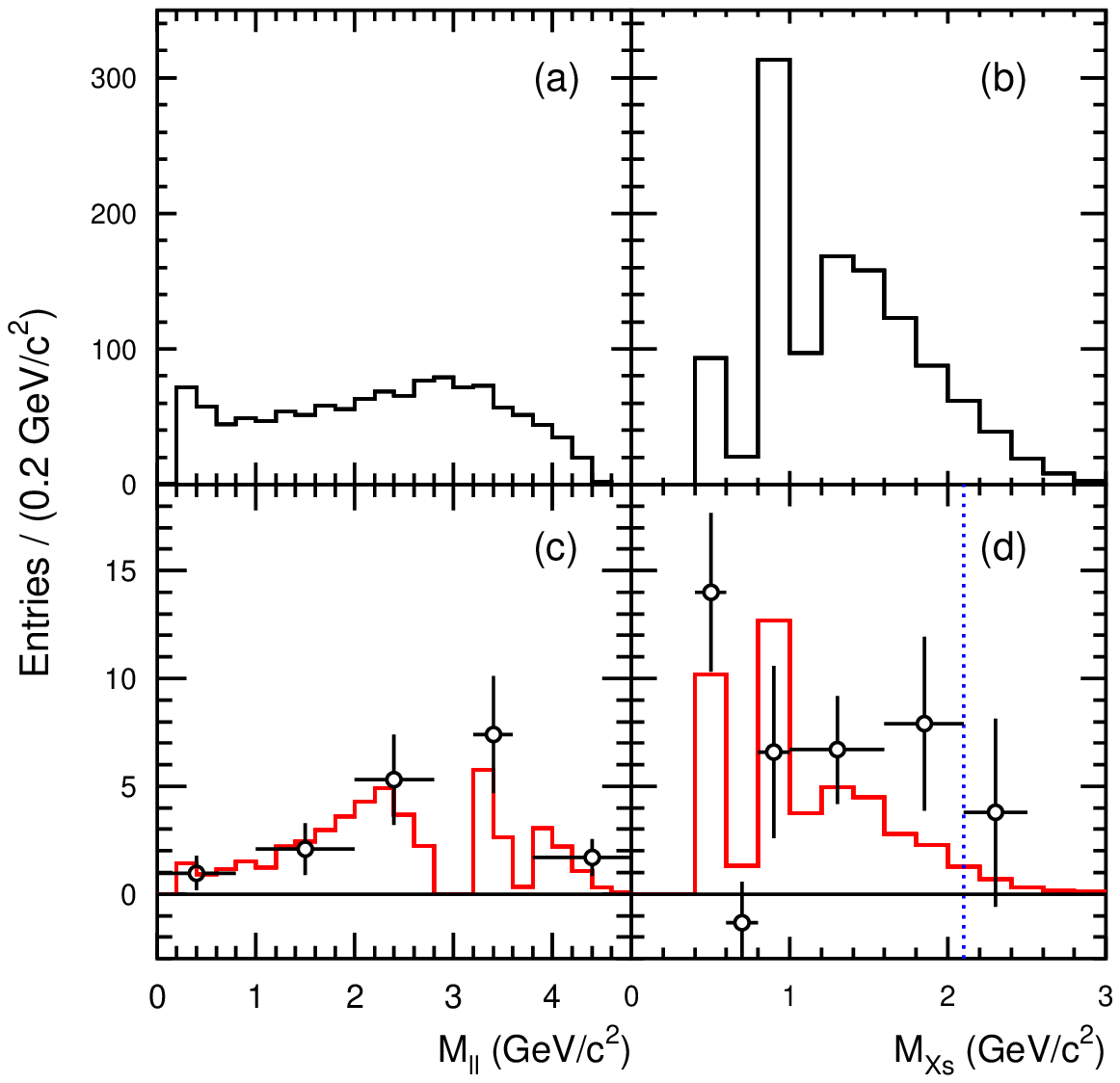}
  \end{center}
 \end{minipage} \\
 \begin{minipage}[b]{0.48\textwidth}
  \caption{Beam-constrained mass distributions for Belle measurement of 
   \bxsll (60\invfb):
   (a) \epem, (b) \mumu, (c) sum of \epem and \mumu, (d) $e^\pm\mu^\mp$.
   The solid lines are the results of unbinned maximum likelihood fits,
   while the dashed lines show the background contributions in these fits.
   \label{fig:xsll_Mbc}}
 \end{minipage}
 \begin{minipage}[b]{0.03\textwidth}
  \mbox{}
 \end{minipage}
 \begin{minipage}[b]{0.48\textwidth}
  \caption{Standard Model expectations for the dilepton (left) and
   hadronic (right) mass distributions of signal events, before (top) and 
   after (bottom) all selection cuts.  Histograms are normalized to the
   predicted branching fraction.  The data points show Belle results, with
   statistical errors.  
   \label{fig:xsll_masses}}
 \end{minipage}
\end{figure}

Topological and \de (Eq.~\ref{eq:de}) cuts are used to reduce substantial 
backgrounds from continuum events and from random combinations of leptons 
from semileptonic \B decays; $M_{X_s} < 2.1\gevcc$ avoids a high-background
region.  Charmonium backgrounds (from $\B\to\jpsi (\psi') X_s$ with
leptonic decay of the \jpsi or $\psi'$) are vetoed using the dilepton mass.
After cuts, the $M_{bc}$ (Eq.~\ref{eq:mes}) spectra are fitted to a 
Gaussian signal plus an
ARGUS background shape\,\cite{bib:argus}, as shown in 
Figure~\ref{fig:xsll_Mbc}.  A signal is observed in the \epem and \mumu
channels, but is absent (as expected) for $e^\pm \mu^\mp$.
Finally, a small background from $X_s\pi^+\pi^-$ with double pion
misidentification (2.6 events for \mumu) is subtracted from the fit signal.

Combining the \epem and \mumu results, Belle finds 57.4 net signal events,
with a significance of $5.4\sigma$.  Averaging the two modes, they find
$\BR(\bxsll) = [6.1 \pm 1.4 (\mathrm{stat}) {}^{+1.4}_{-1.1}
   (\mathrm{syst})] \times 10^{-6} $, where the systematic uncertainty
includes $\approx 0.8$ from model-dependence 
in \kaon and \Kstar fractions, the \mxs model, and quark fragmentation.
This result is in agreement with the Standard Model expectations.
The distributions of events in dilepton and hadronic masses, shown in
Figure~\ref{fig:xsll_masses}, are also compatible with expectations.

\section{A Search for $\mathbf{\Bm\to K^-\nu\nub}$}\label{sec:knunu}

The process $\Bm\to\knunu$ occurs in the SM via the same lowest-order 
diagrams as \bxsll (Fig.~\ref{fig:xll_diag}), but only \Z and \Wpm, not 
$\gamma$, can contribute, \ie, there are no long-range effects.  As a
consequence, the SM computation is relatively clean for the
corresponding \emph{inclusive} process.  As usual, 
SM predictions\cite{bib:Buchalla,bib:Faessler} for the 
\emph{exclusive} process $\Bm\to\knunu$ have additional hadronization 
uncertainties.  Summing over three neutrino flavors, Buchalla 
\etal\cite{bib:Buchalla} predict 
$\BR(\Bm\to\knunu) = (3.8{\,}^{+1.2}_{-0.6})\times 10^{-6}$.
Models with altered Flavor-Changing-Neutral-Current \Z couplings
(\forex, SUSY or dynamical Higgs) can increase this by up to a
factor of five.  However, the measurement of \bxsll (section~\ref{sec:bxsll})
already provides tighter constraints.

The lowest published limit, from CLEO\,\cite{bib:cleo_knunu}, is 
$2.4\times 10^{-4}$.  Last year \babar\,\cite{bib:babar_knunu_semi}  set a 
preliminary limit of $9.4\times 10^{-5}$, based on 50.7\invfb of data and 
using semi-leptonic tagging of the other \B.  

The new \babar\ analysis\,\cite{bib:babar_knunu} to be described here, based 
on 80.1\invfb, is entirely independent statistically.  It relies on an 
orthogonal tagging sample, using full reconstruction of the other \B via
$\Bp\to\Dzb X^+$, where $\X^+$ is composed of up to three charged mesons 
($\pi$ or \kaon) and up to two \piz{}s.   The \Dzb is reconstructed in 
$K^+\pim$, $K^+\pim\piz$ or $K^+\pim\pip\pim$.

\begin{figure}[htb]
 \begin{center}
  \includegraphics[width=0.65\textwidth,clip]{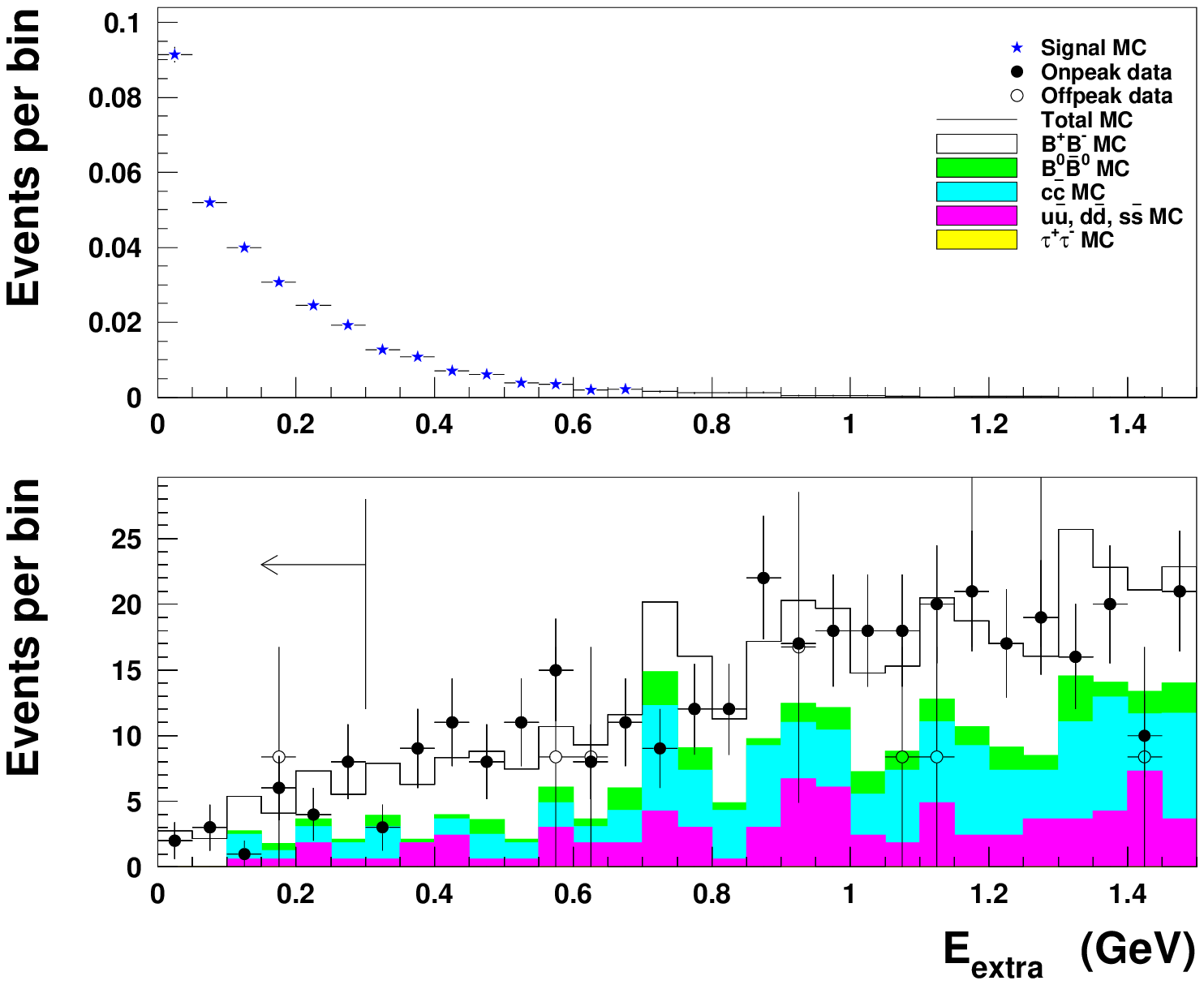}
  \vspace{-0.1in}
  \caption{Simulated (Monte Carlo) distributions of ``extra'' neutral
   energy (see text) for the \babar\ 
   $\Bm\to\knunu$ search.  The top plot is for signal, assuming 
   $\BR(\Bm\to\knunu) = 4.0\times 10^{-6}$, while
   the bottom plot shows cumulative continuum and generic \BB backgrounds.  
   Both are scaled to the luminosity of the data, and are after the
   signal-side requirements of an identified kaon and no other tracks,
   but no other signal-side cuts.
   \label{fig:knunu_eneut}}
 \end{center}
\end{figure}

Continuum backgrounds are reduced by topological cuts.  \de and \mes 
for the \emph{tag} candidates (see Eqs.~\ref{eq:de} 
and~\ref{eq:mes}), are required to be between -1.8 and 0.6\gev, and
above 5.2\gevcc, respectively.  Any ambiguity is resolved by choosing
the candidate with \de closest to 0.  The resulting simulated \mes
spectra show a clear peak only for \BpBm events.  Events
with $5.272 < \mes < 5.288\gevcc$ are then used to search for the
signal, using everything not included in the reconstructed tag.  
Signal-side requirements are:  only one charged track (opposite charge to 
the tag), identified as a kaon; kaon momentum at least 1.5\gevc in the CM; 
the total missing momentum vector pointing into the detector acceptance;
no pair of calorimeter clusters consistent with a \piz mass; and total
extra neutral CM-frame energy $E_{extra} < 300\mev$.  The latter is a 
particularly powerful cut, as can be seen in Figure~\ref{fig:knunu_eneut}.
The overall efficiency for signal events is $(0.0458\pm 0.0046)\%$.

A combinatoric background of $1.0\pm 0.4$ events is estimated by
extrapolating from an \mes sideband, while a peaking background of
$1.7\pm 0.6$ events is taken from \BpBm Monte Carlo simulation.
The total background of $2.7\pm 0.7$ compares to 3 data events
in the unblinded signal region.

By using a large number of parameterized (``toy'') Monte Carlo 
experiments for each signal hypothesis, a preliminary upper limit
of  $\BR(\Bm\to\Km\nu\nu) < 1.05\times 10^{-4}$ (90\% CL) is found.
The efficiency quoted above is based on the nominal $K$ momentum
spectrum given by Buchalla \etal\cite{bib:Buchalla}  Using their
extreme spectra scales the upper limit to 1.10 or 1.02, while using
the curve in Faessler \etal\cite{bib:Faessler} results in 0.95
(all $\times 10^{-4}$).  Combining the nominal result with the earlier, 
independent, \babar\ 
measurement\,\cite{bib:babar_knunu_semi} results in 
$\BR(\Bm\to\Km\nu\nu) < 7.0\times 10^{-5}$.

\section{Searches for $\mathbf{\Bp\to\tau^+\nut}$}\label{sec:taunu}

Purely leptonic \Bp decays in the SM occur via an s-channel \Wp and
have cleanly computed branching fractions:
\be
 \BR(\Bp\to\ell^+\nu) = \frac{G_F^2 m_{\B} m_{\ell}^2} {8\pi}
 \biggl( 1- \frac{m_{\ell}^{2}}{m_{\B}^2} \biggr)^2 f_{\B}^2 
 \Vub^2 \tau_{\B}\ \ .
\ee
From current PDG values\,\cite{bib:PDG}, using lattice gauge theory for
$f_B$, $\BR(\Bp\to\taunu) \approx 7.5\times 10^{-5}$.  (The \munu final
state would be much easier to measure, but its prediction is
$\approx 250\times$ smaller, a result of ``helicity suppression'', \ie, 
the $\ell^+$ is produced left-handed.)  Depending on the state of external
knowledge of $f_B\Vub$, a measurement could either provide this product
(within the SM) or provide evidence for new physics (\forex, a charged Higgs,
or leptoquark exchange).  The most stringent published limit is from 
L3\,\cite{bib:l3_taunu}, $ < 5.7\times 10^{-4}$ at 90\% CL.

\babar\ now has two new independent preliminary results,
both using 81.9\invfb of data: 
\begin{itemize}
 \item[$\bullet$] One\,\cite{bib:babar_taunu_semi} uses semi-leptonic 
  \B decays as tags, and considers only the leptonic decay modes of the 
  \taup, \ie, $\ep\nue\nutb$ and $\mup\num\nutb$.
 \item[$\bullet$] The other\,\cite{bib:babar_taunu_reco} uses 
  fully-reconstructed \B{}s for tagging, and considers five \taup modes,
  adding $\pip\nutb$, $\pip\piz\nutb$, and $\pip\pim\pip\nutb$.
\end{itemize}

Because the basics of the second method are similar to those of the \babar\ 
\knunu analysis (sec.~\ref{sec:knunu}), this summary focuses on the quite
different semi-leptonic-tag method.  The \taup is observed via its single
charged lepton; nothing is missing except neutrinos.  Hence any other
tracks or neutral energy must arise from the other (tagging) \Bm.  Keeping
only events with a total event charge of 0, this ``tag'' is reconstructed
as $\Bm\to\Dz\ellm\nub_{\ell} X$, beginning with a reconstructed \Dz in a
$\Km\pip$, $\Km\pip\piz$, $\Km\pip\pip\pim$ or $\KS\pip\pim$ mode, and
vertexing this with a high-momentum lepton.  If there are multiple $D\ell$
candidates, the best is selected using the \Dz mass.  If $X$ is assumed to
be null, there is enough information to compute the cosine of the angle
between the parent \B and the $D\ell$.  If the assumption is wrong, this
cosine can take on values beyond $\pm 1$; a generous cut allows for
feeddown from higher-mass neutral $D$ states (\forex, from 
$\Dstarz\to\Dz\piz$) while still suppressing background.

\begin{figure}[htb]
 \begin{minipage}[c]{0.48\textwidth}
  \begin{center}
   \includegraphics[width=\textwidth,clip]{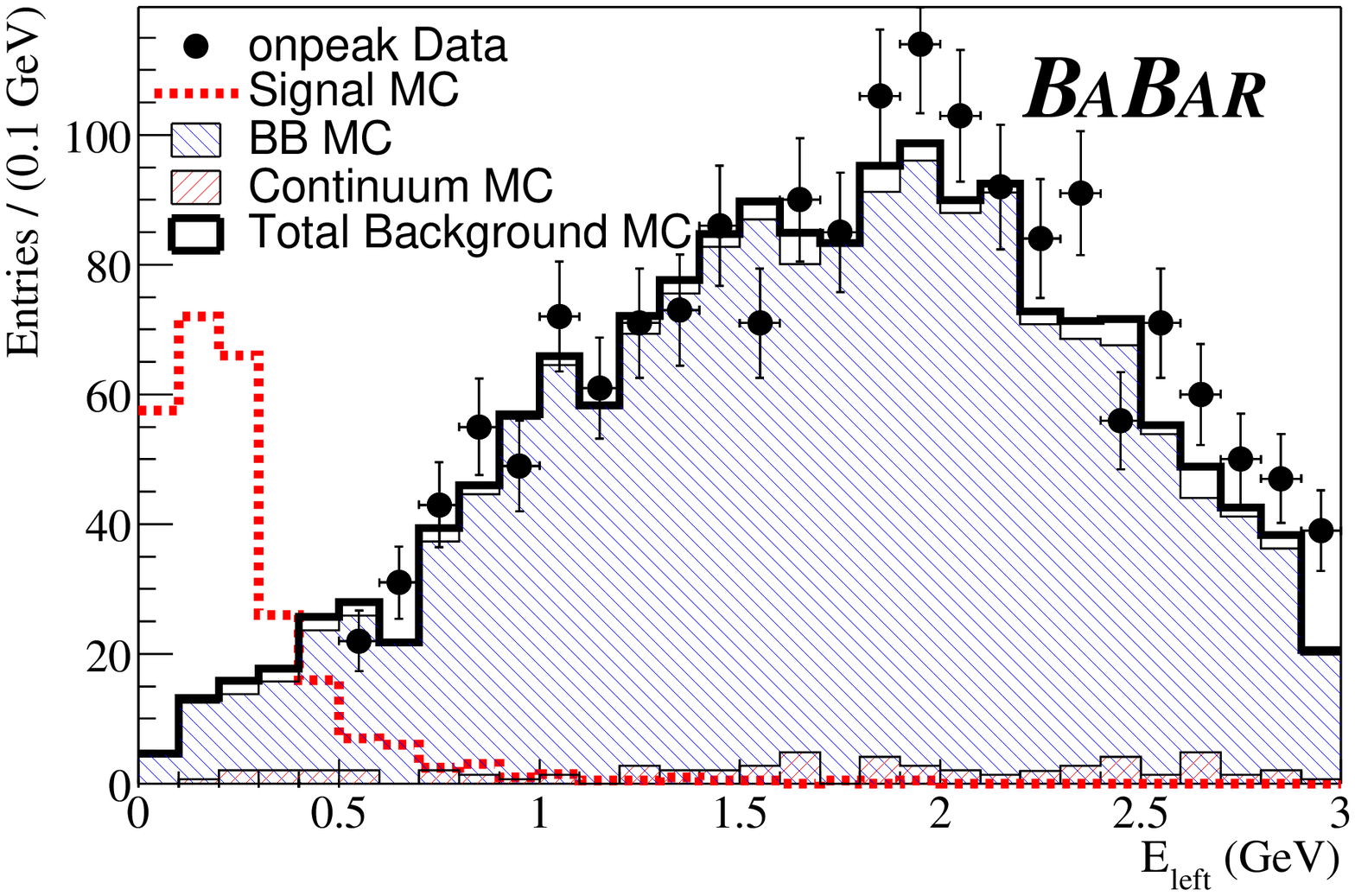}
  \end{center}
 \end{minipage}
 \begin{minipage}[c]{0.03\textwidth}
  \mbox{}
 \end{minipage}
 \begin{minipage}[c]{0.48\textwidth}
  \begin{center}
   \includegraphics[width=\textwidth,clip]{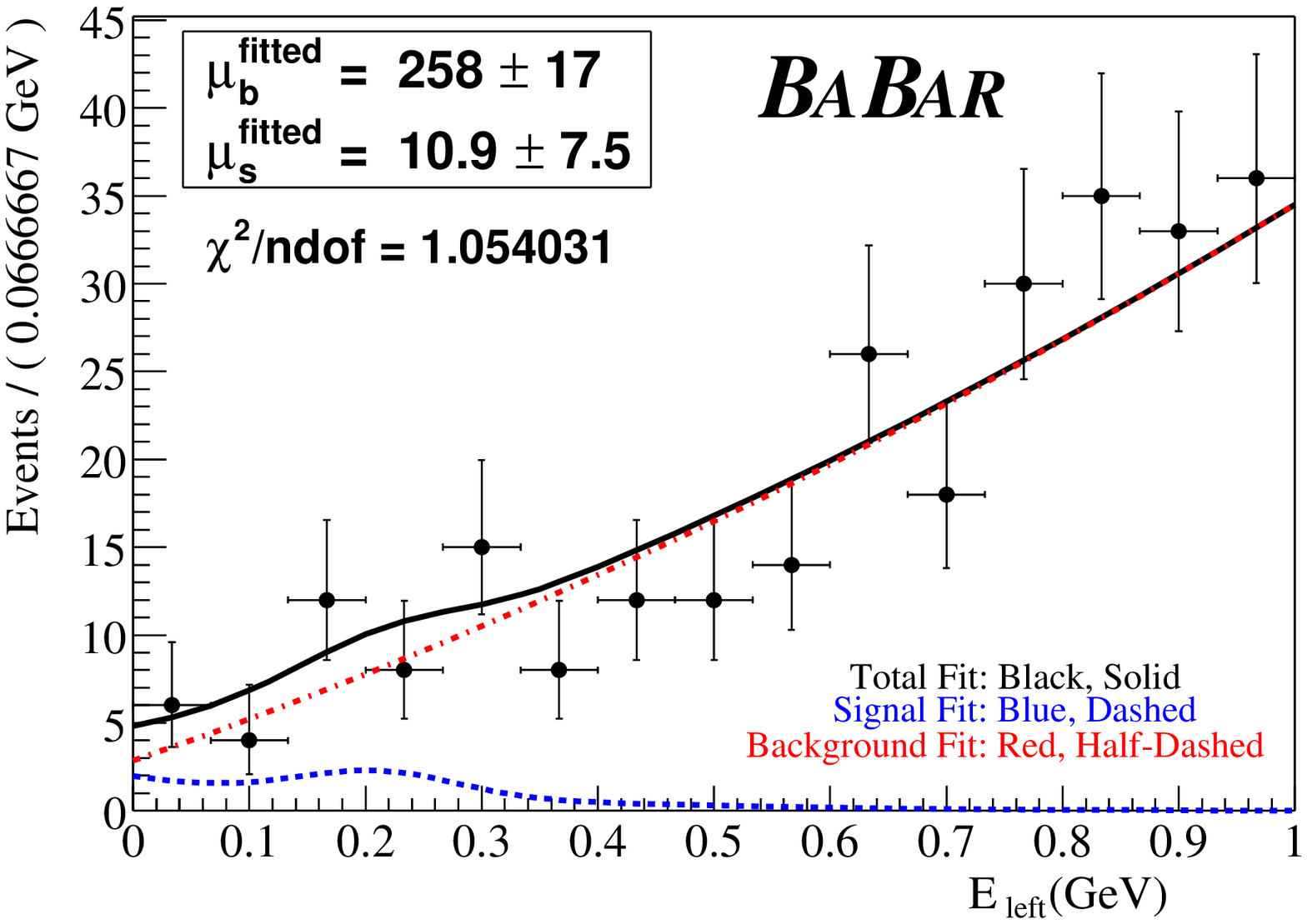} 
  \end{center}
 \end{minipage} \\[-0.15in]
 \begin{minipage}[b]{0.48\textwidth}
  \caption{$E_{left}$ after all cuts for \babar\ measurement of 
   \Bp\to\taunu using semi-leptonic tags (preliminary).  The signal MC has 
   arbitrary normalization.  (Data points below 0.5\mev were kept 
   hidden at this stage, until systematic studies and fitting and
   and limit-setting procedures were settled.)
   \label{fig:taunu_eleft}}
 \end{minipage}
 \begin{minipage}[b]{0.03\textwidth}
  \mbox{}
 \end{minipage}
 \begin{minipage}[b]{0.48\textwidth}
  \caption{$E_{left}$ distribution of data (no longer hidden) after all
   cuts for \babar\ measurement of \Bp\to\taunu using semi-leptonic tags 
   (preliminary).  Curves show the maximum likelihood fit, along with its 
   signal and background components.  The quoted $\mu$ values are the
   fitted numbers of events.
   \label{fig:taunu_fit}}
 \end{minipage}
\end{figure}

Anything remaining after tag reconstruction is assigned to the ``signal
side''.  There must be a single charged track, identified as \ep or \mup.
Additional cuts reject \tautau continuum events.  The \emph{total 
``signal-side'' neutral energy} $E_{left}$  is the signal-defining quantity
in this analysis.  (For a real \taunu event, non-zero $E_{left}$ can arise
from beam backgrounds or from unused particles contributing to $X$ in the 
decay of the tag \B.)  Figure~\ref{fig:taunu_eleft} shows distributions for
signal and background MC simulation, along with data in the background
region.  An extended maximum likelihood fit is then used to fit the data
to signal plus background between 0 and 1\gev; the background simulation
predicts 269 events in this fit region for the \babar\ data sample.

The fitted signal is converted to a 90\% CL upper limit using a version
of the approach\,\cite{bib:Read} 
devised for Higgs searches at LEP, in this case implemented
via many ``toy'' MC experiments for each signal hypothesis.  The median
upper limit for background-only toy-MC is 15.4 events, which can be 
regarded as the \emph{sensistivity} of this measurement.  Two independent
dataset-sized backround-only \emph{full} simulations were run in order to
study systematic effects.  The ``Higgs procedure'' applied to these
yielded upper limits of 13.3 and 19.4 events, close to the sensitivity.
These full simulations were also used to study possible variations in
the signal and background $E_{left}$ distribution functions, with the 
conclusion that the default analysis is conservative for the purpose
of setting an upper limit.

Figure~\ref{fig:taunu_fit} shows the results of the $E_{left}$ fit.
The 10.9 fitted signal events translate to a 90\% CL limit of 22.8
(well within the spread of background-only toy-MC outputs).  Taking into
account the signal efficiency and its systematic uncertainty (determined
using data control samples), the result is
$\BR(\Bp\to\taup\nut) < 4.9\times 10^{-4}$ (90\% CL, preliminary). 
The independent \babar\ measurement using fully-reconstructed tags resulted 
in $\BR(\Bp\to\taup\nut) < 7.7\times 10^{-4}$ (90\% CL, preliminary).
The combined \babar\ upper limit is then
$\BR(\Bp\to\taup\nut) < 4.1\times 10^{-4}$ (90\% CL, preliminary).
This is still a factor of 5.5 above the SM prediction, but is somewhat lower
than the previous best limit.

\section{Conclusions}\label{sec:conclusions}

Rare \B decays have the potential for exhibiting physics beyond the SM,
but no deviation has yet been demonstrated.  
Limits on several unseen exclusive decay modes (\rhog, \omegag, \knunu,
\taunu) have come down significantly since last year's Moriond
EW conference, and it is important to push them (and others, like \munu)
further, since non-SM
effects could in some cases enter at levels significantly above the SM
predictions.  The limits for $\B\to\rho(\omega)\gamma$ 
(section~\ref{sec:rho}) are now only a factor of two above the \emph{low}
end of SM predictions, so we may be close to observing a $\b\to\d\gamma$
signal for the first time.

Meanwhile, the first observation of inclusive \bxsll (section~\ref{sec:bxsll})
opens up a rich new area of investigation.  The exclusive channels \kll
(section~\ref{sec:kll}, with the $K$ mode established, the \Kstar mode
nearly so) are already offering some tantalizing SM comparisons, albeit
with the usual caveat that predictions for exclusive modes are considerably
more uncertain than those for inclusive modes.  Measurements of
inclusive \xsg (section~\ref{sec:bxsg}) are moving toward useful precision
on the \eg spectrum and its moments, which are linked theoretically
to semileptonic decays; while the branching fraction values already show
that any SM deviation in this sector (more limited than that accessible with
\xsll) must be fairly small.  Finally, the first published measurement on
radiative decay to a higher \Kstar resonance (section~\ref{sec:higher})
marks the start of an exploration of the detailed composition of the 
inclusive \bxsg process

Both \babar\ and Belle are continually updating results to new data and
improving analysis techniques, and we can expect continued progress in
all of the above areas and more.


\section*{Acknowledgments}
The author thanks the physicists of the Belle collaboration and the \babar\
collaboration (of which he is a member) for providing the figures and
detailed explanations needed to prepare this report.  Its preparation
was supported in part by the U.S. Department of Energy, 
\#DE-FG03-92ER40689.


\begin{thebibliography}{99}
\bibitem{bib:Gambino} P.~Gambino and M~Misiak, \Journal{\NPB}{611}{338}{2001}.

\bibitem{bib:Buras} A.J.~Buras and M.~Misiak,
 \Journal{\em{Acta Phys. Polon.} B}{33}{2597}{2002}.

\bibitem{bib:PDG} Particle Data Group, \Journal{\PRD}{66}{010001}{2002}.

\bibitem{bib:Ali_Parkhomenko} A.~Ali and A.Y.~Parkhomenko, 
 \Journal{\em{Eur. Phys. J.} C}{23}{89}{2002}.

\bibitem{bib:Bosch} S.W.~Bosch and G.~Buchalla, \Journal{\NPB}{621}{459}{2002}.

\bibitem{bib:babar_rho} B.~Aubert \etal, \babar\ Collaboration,
 SLAC-PUB-9938, BABAR-PUB-03/008 (2003).

\bibitem{bib:belle_rho} K.~Abe \etal, Belle Collaboration,
 BELLE-CONF-0240 (2002); results updated via private communication, 
 T.~Browder and M.~Nakao.

\bibitem{bib:BABAR} B.~Aubert \etal, \babar\ Collaboration,
 \Journal{\NIMA}{479}{1}{2002}.

\bibitem{bib:argus} H.~Albrecht \etal, ARGUS Collaboration,
 \Journal{\ZPC}{48}{543}{1990}.

\bibitem{bib:Cousins} R.~Cousins and V.~Highland, 
 \Journal{\NIMA}{320}{331}{1992}.

\bibitem{bib:cleo_rho} T.E.~Coan \etal, CLEO Collaboration,
 \Journal{\PRL}{84}{5283}{2000}.

\bibitem{bib:higher} S.~Nishida \etal, Belle Collaboration,
 \Journal{\PRL}{89}{231801}{2002}.

\bibitem{bib:Veseli} S.~Veseli and M.G.~Olsson,
 \Journal{\PLB}{367}{309}{1996}.

\bibitem{bib:Kagan} A.L.~Kagan and M.~Neubert,
 \Journal{\em{Eur. Phys. J.} C}{7}{5}{1999}.

\bibitem{bib:inclbsg_babar} B.~Aubert \etal, \babar\ Collaboration,
 SLAC-PUB-9301, BABAR-CONF-02/26 (2002).

\bibitem{bib:semibsg_babar} B.~Aubert \etal, \babar\ Collaboration,
 SLAC-PUB-9308, BABAR-CONF-02/25 (2002).

\bibitem{bib:Bigi} I.~Bigi and N.~Uraltsev,
 \Journal{\em{Int. J. Mod. Phys.} A}{17}{4709}{2002}.

\bibitem{bib:Ligeti} Z.~Ligeti, M.E.~Luke, A.V.~Manohar and M.B.~Wise,
 \Journal{\PRD}{60}{034019}{1999}, plus private communication from
 Z.~Ligeti and M.B.~Wise.

\bibitem{bib:firstKll} K.~Abe \etal, Belle Collaboration,
 \Journal{\PRL}{88}{021802}{2002}.

\bibitem{bib:belle_Kll} K.~Abe \etal, Belle Collaboration, 
 BELLE-CONF-0241 (2002).

\bibitem{bib:babar_Kll} B.~Aubert \etal, \babar\ Collaboration,
 SLAC-PUB-9323, BABAR-CONF-02/023 (2002).

\bibitem{bib:Ali_2002} A.~Ali, E.~Lunghi, C.~Greub and G.~Hiller,
 \Journal{\PRD}{66}{034002}{2002}.

\bibitem{bib:Ali_2000} A.~Ali, P.~Ball, L.T.~Handoko and G.~Hiller, 
 \Journal{\PRD}{61}{074024}{2000}.

\bibitem{bib:belle_xsll} J.~Kaneko \etal, Belle Collaboration,
 \Journal{\PRL}{90}{021801}{2003}.

\bibitem{bib:Ali_1997} A.~Ali, G.~Hiller, L.T.~Handoko and T.~Morozumi, 
 \Journal{\PRD}{55}{4105}{1997}.

\bibitem{bib:Buchalla} G.~Buchalla, G.~Hiller and G.~Isidori,
 \Journal{\PRD}{63}{014015}{2001}.

\bibitem{bib:Faessler} A.~Faessler \etal,
 \Journal{\em{Eur. Phys. J.} C}{4}{18}{2002}.

\bibitem{bib:cleo_knunu} T.E.~Browder \etal, CLEO Collaboration,
 \Journal{\PRL}{86}{2950}{2001}.

\bibitem{bib:babar_knunu_semi} B.~Aubert \etal, \babar\ Collaboration,
  SLAC-PUB-9309, BABAR-CONF-02/027 (2002).

\bibitem{bib:babar_knunu} B.~Aubert \etal, \babar\ Collaboration,
 SLAC-PUB-9710, BABAR-CONF-03/006 (2003).

\bibitem{bib:l3_taunu} M.~Acciari \etal, L3 Collaboration,
 \Journal{\PLB}{396}{327}{1997}.

\bibitem{bib:babar_taunu_semi} B.~Aubert \etal, \babar\ Collaboration,
 SLAC-PUB-9688, BABAR-CONF-03/005 (2003).

\bibitem{bib:babar_taunu_reco} B.~Aubert \etal, \babar\ Collaboration,
 SLAC-PUB-9716, BABAR-CONF-03/004 (2003).

\bibitem{bib:Read} A.L.~Read, {\em{First Workshop on Confidence Limits}}, 
 CERN-2000-005, 81 (2000).

\end{thebibliography}
\end{document}